\documentclass[%
 reprint,
groupedaddress,
 amsmath,amssymb,
 aps,
]{revtex4-2}

\usepackage{graphicx}% Include figure files
\usepackage{dcolumn}% Align table columns on decimal point
\usepackage{bm}% bold math
\usepackage{color}
\usepackage{amsmath}
\usepackage{hyperref}% add hypertext capabilities
\begin{document}

\title{Two-dimensional quantum Griffith singularity in three-dimensional ZrN$_x$
 superconducting films}% Force line breaks with \\

\author{Zi-Yan Han}\thanks{These authors contributed equally to this work.}
\affiliation{Tianjin Key Laboratory of Low Dimensional Materials Physics and
Preparing Technology, Department of Physics, Tianjin University, Tianjin 300354,
China}
\author{Li-Min Yu}\thanks{These authors  contributed equally to this work.}
\affiliation{Tianjin Key Laboratory of Low Dimensional Materials Physics and
Preparing Technology, Department of Physics, Tianjin University, Tianjin 300354,
China}
\author{Yu-Cheng Cong}
\affiliation{Tianjin Key Laboratory of Low Dimensional Materials Physics and
Preparing Technology, Department of Physics, Tianjin University, Tianjin 300354,
China}
\author{Yang Yang}
\affiliation{Tianjin Key Laboratory of Low Dimensional Materials Physics and
Preparing Technology, Department of Physics, Tianjin University, Tianjin 300354,
China}
\author{Zhi-Xiang Sun}
\affiliation{Tianjin Key Laboratory of Low Dimensional Materials Physics and
Preparing Technology, Department of Physics, Tianjin University, Tianjin 300354,
China}
\author{Zhi-Qing Li}
\email[Corresponding author, e-mail: ]{zhiqingli@tju.edu.cn}
\affiliation{Tianjin Key Laboratory of Low Dimensional Materials Physics and
Preparing Technology, Department of Physics, Tianjin University, Tianjin 300354,
China}
\date{\today}% It is always \today, today,

\begin{abstract}
We report the experimental observation of two-dimensional (2D) quantum Griffiths singularity (QGS) in $\sim$200-nm-thick epitaxial ZrN$_x$ superconducting films. The films possess a rock-salt structure and are three-dimensional (3D) with respect to superconductivity. For each film with $x \gtrsim 1.30$, the low-temperature magnetoresistance isotherms under fields perpendicular and parallel to the film plane cross over at a broad magnetic field range independently rather than at a single crossing point. Despite the macroscopic 3D nature of the superconductivity, the magnetoresistance isotherms at selected adjacent temperatures follow the theoretical prediction of power-law scaling for 2D superconducting systems, rather than that for 3D systems. The effective critical exponent $z\nu$, obtained by analyzing the magnetoresistance isotherms using the 2D power-law scaling, increases with decreasing temperature and diverges as the quantum phase transition is approached. In addition, the resistivity data near the superconductor-insulator or superconductor-metal transitions obey an activated scaling form that describes the quantum phase transition of 2D superconducting systems governed by an infinite-randomness critical point. The QGS in the ZrN$_x$ films is attributed to quenched disorder induced by intrinsic defects, such as Zr vacancies and N interstitials, which creates spatially inhomogeneous superconducting rare regions. The dynamics of these rare regions, which may exhibit effective 2D characteristics near the quantum critical point, dominate the transport properties of the system near the quantum phase transition. Our results provide compelling evidence for the existence of QGS in 3D superconductors and highlight the crucial role of disorder-induced inhomogeneity in determining the critical behavior of quantum phase transitions.
\end{abstract}

\maketitle

%\tableofcontents

\section{Introduction}\label{secI}

Over the past decade, quantum Griffiths singularity (QGS) in two-dimensional (2D) superconductors has attracted great attention~\cite{Saito-2016}. In the framework of the ``dirty-boson model", a magnetic field would induce a quantum phase transition (QPT) in 2D superconductors at zero temperature, driving the system to transition from a superconducting state to an insulating (or metallic) state. Meanwhile, in the vicinity of the transition, the magnetic field dependence of resistance curves at different temperatures all cross at one point and obey a power-law scaling form~\cite{Fisher-1989,Fisher-1990}
\begin{equation}\label{Eq-dirtyBoson}
R(B, T)=R_{\rm c} f(\delta T^{-1/z\nu})
\end{equation}
where $z$ is the dynamical critical exponent, $\nu$ is the correlation length exponent, $R_{\rm c}$ is the critical resistance, $\delta=|B-B_{\rm c}|$ is the distance between the magnetic field $B$ and critical magnetic field $B_{\rm c}$, and $f(x)$ is an arbitrary function with $f(0)=1$. The theoretical prediction of the ``dirty-boson model" has been experimentally observed in a variety of 2D superconducting thin films~\cite{Yazdani-1995,Markovic-1998,Mason-1999,Marrache-Kikuchi-2008,Steiner-2008,Breznay-2016}, in which the low-temperature magnetoresistance isotherms cross at one point and obey the aforementioned scaling law.
In 2015, Xing \emph{et al}.~\cite{Xing-2015} reported the emergence of QGS in three-monolayer Ga films during the magnetic field-induced superconductor–metal transition. Near the transition the low-temperature magnetoresistance isotherms of each film do not cross at a single point, but rather in a wide region. Assuming that the magnetoresistance isotherms at three adjacent temperatures cross at one point, they obtained the critical magnetic field $B_{\rm c}$ dependence of the critical exponent $z\nu$. It has been found that $z\nu$ diverges as $B_c \rightarrow B_c^\ast$ or $T \rightarrow 0$\,K, where $B_c^\ast$ is the critical field as  $T\rightarrow 0$\,K. This behavior was taken as evidence for the QGS associated with an infinite-randomness critical point. Thereafter, the existence of QGS has been found in a range of 2D superconducting systems~\cite{Shen-2016,Xing-2017,Saito-2018,Lewellyn-2019,Zhang-2019,Verzhbitskiy-2020,Huang-2021,Liu Y-2021,Jing-2023}.

Theoretically, QGS is also predicted to occur in 3D systems, and it has been experimentally confirmed in a range of 3D magnetic materials for decades~\cite{Andrade-1998,Neto-1998,Ubaid-Kassis-2010,Steppke-2013}. In contrast, the emergence of QGS in 3D superconductors has been reported only recently, specifically in the MgTi$_2$O$_4$ system~\cite{Qi-2024}, the Mo$_{0.8}$Ti$_{0.2}$N$_x$ series~\cite{Wang-2024}, and the CaFe$_{1-x}$Ni$_x$AsF series~\cite{Liu-2025}. Moreover, the characteristics of the QGS in these three systems are not entirely the same. For the 3D MgTi$_2$O$_4$~\cite{Qi-2024} and CaFe$_{1-x}$Ni$_x$AsF~\cite{Liu-2025} superconductors, the critical magnetic field dependence of the critical exponent $z\nu$ obeys the activated scaling law for 3D systems, while for MoTiN films~\cite{Wang-2024}, the critical magnetic field dependence of $z\nu$ obeys the activated scaling law for 2D systems though these films are 3D with respect to the superconductivity. Thus, it is nontrivial to explore the QGS in more 3D superconducting materials and investigate the scaling relations of the critical exponent $z\nu$ with the critical field $B_c$ and the resistance with magnetic field and temperature near the QPT point.

As early as the 1930s, the zirconium nitride (ZrN) with a rock-salt structure was found to exhibit superconductivity below 9.05\,K~\cite{Matthias-1952}. Recently, it is found~\cite{Chen-2023} that ZrN$_x$ films maintain the NaCl structure and retain their superconducting properties within a nitrogen composition range of approximately $0.61 \lesssim x \lesssim 1.33$. In addition, a superconducting dome emerges in the phase diagram of ZrN$_x$ films near the boundary between superconducting and strongly insulating phases. This feature resembles that of copper-based high-temperature superconductors~\cite{Keimer-2015}, despite the superconductivity in ZrN$_x$ films following the Bardeen–Cooper–Schrieffer mechanism~\cite{Bardeen-1957}. On the other hand, point defects such as Zr vacancies and nitrogen interstitials are inevitable in ZrN$_x$ films in the nitrogen-rich region ($x>1$)~\cite{Chen-2023}, similar to other transition metal nitrides. These defects introduce quenched disorder, which could lead to inhomogeneous superconductivity in the films and potentially result in a QGS.  Additionally, ZrN$_x$ films possess relatively low upper critical magnetic fields~\cite{Potjan-2023}. These features could make ZrN$_x$ films a suited system for exploring the characteristics of QGS in 3D supeconducting systems.

In this work, we fabricated a series of $\sim$200-nm-thick ZrN$_x$ films with $x$ ranging from $\sim$1.21 to $\sim$1.50 by finely controlling the nitrogen partial pressure. We systematically investigated their structural and low-temperature electrical transport properties. The results reveal that superconductivity in these ZrN$_x$ films is three-dimensional. Furthermore, QGS emerges in films with $x \gtrsim 1.30$ under both perpendicular and parallel magnetic fields. Interestingly, despite this 3D superconducting nature, the critical magnetic field dependence of $z\nu$ follows the scaling law typically associated with 2D superconductors with QGS. Moreover, resistivity data near the superconductor-insulator (or metal) transition obey an activated scaling form indicative of 2D QPTs governed by an infinite-randomness critical point. The possible origins of these observations are discussed.

\section{Experimental method}\label{secII}
The ZrN$_x$ films were deposited on (100) MgO single crystal substrates by the reactive \emph{rf} sputtering technique. A commercial Zr target with a purity of 99.95\% and a diameter of 60\,mm served as the sputtering source. Prior to deposition, the chamber was evacuated to a base pressure of $\sim$$1.5 \times 10^{-4}$\,Pa. Then the chamber pressure was adjusted to 0.2 Pa by backfilling with a mixture of nitrogen and argon. During the deposition process, the sputtering power applied to the Zr target was maintained at 300\,W, and the substrate temperature was held constant at $\sim$740\,K. To obtain ZrN$_x$ films with different superconducting transition temperatures, the nitrogen-to-argon volume ratio was systematically adjusted in each deposition run. For the films used in this study, the nitrogen-to-argon volume ratios were set to $1:19$, $1:9$, $3:17$, $7:33$, and $1:4$, corresponding to nitrogen partial pressures of approximately $P_{\rm N_2}\simeq 5.0\%$, 10.0\%, 15.0\%, 17.5\%, and 20.0\%, respectively.

The thicknesses of the films were controlled to $\sim$200\,nm by adjusting the deposition time and were finally determined by a surface profiler (Dektak, 6M). The crystal structure and phase identification were characterized by x-ray diffraction (XRD), using both $\theta$-2$\theta$ and $\phi$ scans. The composition of each sample was determined by the energy-dispersive x-ray spectroscopy analysis (EDS; EDAX, model Apollo X). The resistivity versus temperature and magnetic field were measured using the standard four-probe method in a physical property measurement system (PPMS-6000, Quantum Design) equipped with a $^3$He refrigerator. The magnetic field was applied perpendicular and parallel to the film plane in separate measurements. For clarity, we use $\perp$ ($\parallel$) as a superscript or subscript to denote quantities measured in the out-of-plane (in-plane) configuration. Hall-bar shaped films (1.0-mm wide and 10.0-mm long, and the distance between the two voltage electrodes is 3\,mm), defined by mechanical masks, were used for transport measurement. To obtain good contact, Ti/Au electrodes were deposited on the films.

\begin{figure}[htp]
\includegraphics [scale=1.08]{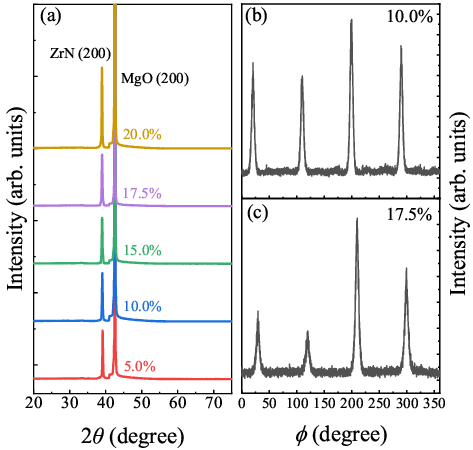}% Here is how to import EPS art
\caption{\label{Fig-XRD}(a) XRD $\theta$-2$\theta$ patterns for ZrN$_x$ films deposited at different nitrogen partial pressures. (b) and (c) are $\phi$-scan spectra of (220) plane for the $x \simeq 1.30$ and $x \simeq 1.42$ films, respectively.}
\end{figure}

\section{Results and Discussions}\label{secIII}
\subsection{Structure and fundamental transport properties}
The EDX results indicate that the atomic ratio of nitrogen to zirconium ($x$) in the ZrN$_x$ films increases from $\sim$1.21 to $\sim$1.50 as the nitrogen partial pressure is enhanced from 5.0\% to 20.0\%. The specific value of $x$ for each film is provided in Table~\ref{tab-1}. The XRD $\theta$-2$\theta$ patterns of the ZrN$_x$ films deposited at different nitrogen partial pressures are shown in Figure~\ref{Fig-XRD}(a). The strong peak at $42.73^{\circ}$ corresponds to the MgO (200) diffraction, which originates from the MgO substrate. For each film, in addition to the MgO (200) diffraction peak, only the (200) peak of fcc-ZrN is observed. As $x$ varies from 1.21 to 1.50, the position of the ZrN (200) peak shifts from $38.94^{\circ}$ to $39.15^{\circ}$, indicating that the lattice constant $a$ of the films ranges from 4.604 to 4.627~\AA. These values are in agreement with those reported in previous studies~\cite{Mei-2013}. The EDX and XRD $\theta$-2$\theta$ scan results demonstrate that ZrN$_x$ films retain the rock-salt structure even when the nitrogen-to-zirconium ratio ($x$) reaches $\sim$$1.50$, which is qualitatively consistent with the findings of Chen \emph{et al}.~\cite{Chen-2023}. The $\phi$-scan spectra of the (220) plane for $x\simeq1.30$ and $x\simeq1.42$ films are presented in Fig.~\ref{Fig-XRD}(b) and \ref{Fig-XRD}(c), respectively. The $\phi$-scan spectra for the $x\simeq1.21$, 1.35, and 1.50 films exhibit similar features to those in Fig.~\ref{Fig-XRD}. In each $\phi$-scan spectrum, four uniformly distributed diffraction peaks are clearly observed, indicating the films are epitaxially grown on the substrates.

\begin{table*}%[htbp]
\caption{
Some parameters of the ZrN$_x$ films deposited at different N$_2$ partial pressures $P_{\rm N_2}$. Here $\rho (300\,{\rm K})$ is the resistivity at 300\,K, $T_{\rm c}$ is the superconducting transition temperature, $n$ is the carrier concentration at 10\,K, and $k_{\rm F}\ell$ is the Ioffe-Regel parameter at 10\,K; $B_{\rm c2}^{\perp}(0)$ and $B_{\rm c2}^{\parallel}(0)$ are the out-of-plane and in-plane upper critical magnetic fields at 0\,K, respectively; $\xi_\perp(0)$ and $\xi_\parallel(0)$ are the out-of-plane and in-plane Ginzburg-Landau coherence length at 0\,K, respectively.}\label{tab-1}
\begin{ruledtabular}
\begin{tabular}{ccccccccccc}
Film    & $P_{\rm N_2}$ & $x$ & $\rho (300\,{\rm K})$  & $T_{\rm c}$  & $n$ & $k_{\rm F}\ell$ & $B_{\rm c2}^{\perp}(0)$ & $B_{\rm c2}^{\parallel}(0)$ & $\xi_{\perp}(0)$& $\xi_{\parallel}(0)$ \\
No. &  (\%) &     & (m$\Omega$\,cm) & (K)    & ($10^{23}\,{\rm cm}^{-3}$) &   & (T)             & (T) &(nm) &(nm)  \\
\hline
1 &  5.0  & 1.21 & 0.43 & 4.66 &  1.41 & 1.99 &      &      &      &    \\
2 &  10.0 & 1.30 & 0.89 & 3.68 &  1.09 & 0.99 & 4.87 & 2.21 & 18.1 & 8.2\\
3 &  15.0 & 1.35 & 1.38 & 3.86 &  0.87 & 0.63 & 5.12 & 2.72 &15.1  &8.0\\
4 &  17.5 & 1.42 & 2.22 & 2.01 &  0.70 & 0.38 & 3.00 & 1.69 &18.6  &10.5\\
5 &  20.0 & 1.50 & 2.64 & 1.73 &  0.65 & 0.31 & 2.19 & 1.60 & 16.8 &12.3\\
\end{tabular}
\end{ruledtabular}
\end{table*}

\begin{figure}[htp]
\includegraphics[scale=1.08]{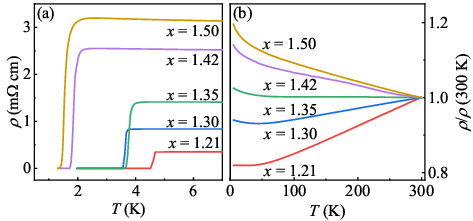}% Here is how to import EPS art
\caption{\label{Fig-RT}(a) Resistivity $\rho$ vs temperature $T$ below 6.5\,K for ZrN{$_x$} films deposited at different nitrogen partial pressures. (b) Normalized resistivity $\rho/\rho(300\,\rm{K})$ vs temperature $T$ from $\sim$5 to 300\,K  for ZrN{$_x$} films deposited at different nitrogen partial pressures.}
\end{figure}

Figure~\ref{Fig-RT}(a) shows the temperature $T$ dependence of the resistivity $\rho$ in the low temperature regime for the films with $1.21 \lesssim x \lesssim 1.50$. The superconducting transition temperature $T_{\rm c}$, defined as the temperature at which the resistance falls to 90\% of its normal-state value $\rho(10\,\rm K)$, decreases with increasing $x$ except for the $x \simeq 1.30$ and $x \simeq 1.35$ films. The value of $T_{\rm c}$ for each film is listed in Tabel~\ref{tab-1}, Figure~\ref{Fig-RT}(b) shows the temperature $T$ dependence of the normalized resistivity $\rho/\rho(300\,\rm K)$ for the above films from 300 down to 5\,K. For these $x\gtrsim 1.35$ films, the normal state resistivity increases with decreasing temperature in the whole temperature range. While for the $x \simeq 1.21$ and 1.30 films, the resistivity initially decreases with decreasing temperature, reaches a minimum at $T_{\rm min}$, and then slightly increases with further decreasing temperature. The values of $T_{\rm min}$ for the $x \simeq 1.21$ and 1.30 films are 27.30 and 45.26\,K, respectively. In Table~\ref{tab-1}, we also give the values of carrier concentration $n$ and the Ioffe-Regel parameter $k_{\rm F} \ell$ ($k_{\rm F}$ is the Fermi wave number, $\ell$ is the electron mean free path)~\cite{Ioffe-1960} measured at 10\,K for all the films. Here, $k_{\rm F} \ell$ is calculated using the free-electron model~\cite{Ashcroft-1976} as $k_{\rm F} \ell = (\hbar/e^2)(3\pi^2)^{1/3}n^{-2/3}\rho^{-1}$, where $\hbar$ is the reduced Planck constant and $e$ is the elementary charge. For the films with $x \gtrsim 1.35$, the Ioffe–Regel parameter is slightly less than 1, whereas the values of $k_{\rm F} \ell$ are $\sim$1.0 and 2.0 for the $x \simeq 1.30$ and 1.21 films, respectively. Together with the $\rho$-$T$ curve characteristics, one can conclude that the three $x \gtrsim 1.35$ films are in the insulator-metal crossover region and the two $x \lesssim 1.30$ films are in a bad metal state. Inspecting Table~\ref{tab-1}, one also sees a monotonic decrease in electron concentration with increasing $x$ at a certain temperature of the normal state. This reduction partially suppresses the Cooper-pair density and, in turn, the superconducting transition temperature~\cite{He-2014}.

\begin{figure}[htbp]
\includegraphics [scale=1.08] {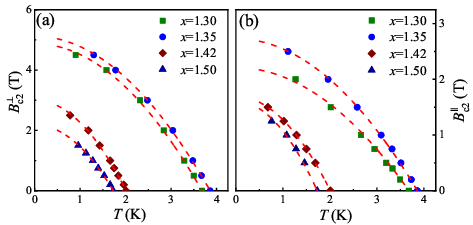}% Here is how to import EPS art
\caption{\label{Fig-Hc2}(a) The out-of-plane upper critical magnetic field $B_{\rm c2}^{\perp}$ vs temperature $T$ for the $x\gtrsim1.30$ films. (b) The in-plane upper critical magnetic field $B_{\rm c2}^{\parallel}$ vs temperature $T$ for the $x\gtrsim1.30$ films.  The dashed curves are the least-squares fits to $B_{\rm c2}^{i}(T)=B_{\rm c2}^{i}(0)[1-(T/T_{\rm c})^2]$ with $i=\perp$ in (a) and $i=\parallel$ in (b).}
\end{figure}

We further measured the temperature-dependent resistivity of the films under both perpendicular and parallel magnetic fields, from which the out-of-plane and in-plane upper critical fields, $B_{\rm c2}^{\perp}$ and $B_{\rm c2}^{\parallel}$, were extracted. Since no QGS was observed in the $x\simeq 1.21$ film (see further remarks below), we restrict our discussion to the results from the other four films. Figure~\ref{Fig-Hc2} (a) and (b) present $B_{\rm c2}^{\perp}$ and $B_{\rm c2}^{\parallel}$ variation with $T$ for the four films, respectively. In 2D superconducting films, the in-plane upper critical field, $B_{\rm c2}^{\parallel}$, is far greater than the perpendicular component, $B_{\rm c2}^{\perp}$, at any given temperature~\cite{Devarakonda-2020,Lu-2015,Saito-2015,Liu C-2021}. In contrast, for the ZrN$_x$ films, $B_{\rm c2}^{\parallel}$ is observed to be lower than $B_{\rm c2}^{\perp}$ at a fixed temperature, as clearly illustrated in Fig.~\ref{Fig-Hc2}. This anomalous behavior indicates that the films do not behave as 2D superconductors but rather as 3D systems exhibiting significant anisotropy. Furthermore, given that the ZrN$_x$ thin films possess a rock-salt structure with the surface normal aligned along the [100] direction, one would expect $B_{\rm c2}^{\parallel}$ and $B_{\rm c2}^{\perp}$ to be nearly identical in the case of 3D thin films. Consequently, the origin of the observed $B_{\rm c2}^{\perp} > B_{\rm c2}^{\parallel}$ relationship in the ZrN$_x$ films warrants further investigation. From Fig.~\ref{Fig-Hc2}, one can see that both $B_{\rm c2}^{\parallel}$ and $B_{\rm c2}^{\perp}$ increase with decreasing temperature over the tested temperature range for each film. The $B_{\rm c2}^{\perp}$-$T$ and $B_{\rm c2}^{\parallel}$-$T$ data are fitted to the empirical formula $B_{\rm c2}^{i}(T)=B_{\rm c2}^{i}(0)[1-(T/T_{\rm c})^2]$~\cite{Ashcroft-1976,Tinkham-1996}, where $i=\perp$ and $\parallel$ represent the perpendicular-field and parallel-field situations, respectively, and $B_{\rm c2}^{i}(0)$ is the upper critical magnetic field at 0\,K. The fitted results are shown by the dashed curves in Fig.~\ref{Fig-Hc2}, which indicate that both the $B_{\rm c2}^{\perp}$-$T$ and $B_{\rm c2}^{\parallel}$-$T$ data are consistent with the empirical formula for each film. The values of $B_{\rm c2}^{\perp}(0)$ and $B_{\rm c2}^{\parallel}(0)$ are summarized in Table~\ref{tab-1}.

\begin{figure}[htbp]
\includegraphics [scale=1.1] {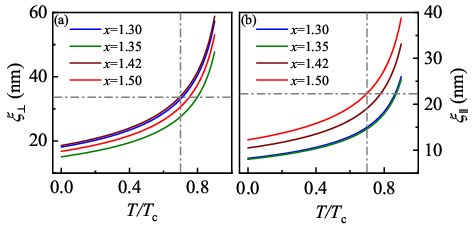}% Here is how to import EPS art
\caption{\label{Fig-Coherence}(a) The out-of-plane GL coherence length $\xi_{\perp}$ vs normalized temperature $T/T_{\rm c}$ for the $x\gtrsim1.30$ films. (b) The in-plane GL coherence length $\xi_{\parallel}$ vs normalized temperature $T/T_{\rm c}$ for the $x\gtrsim1.30$ films.}
\end{figure}

Now we estimate the superconducting coherence length of these films. For an anisotropic 3D superconducting film whose upper critical field depends on the orientation of the applied magnetic field relative to the film surface, the relation between the upper critical field and the Ginzburg-Landau (GL) coherence length can be expressed as~\cite{Tinkham-1996},
\begin{equation}\label{Eq}
 B_{c2}^{\perp}(0) = \frac{\Phi_0}{2\pi\xi_{\parallel}^2(0)},
\end{equation}
and
\begin{equation}\label{Eq}
 B_{c2}^{\parallel}(0) = \frac{\Phi_0}{2\pi\xi_{\perp}(0)\xi_{\parallel}(0)},
\end{equation}
where $\xi_i(0)$  ($i =\perp$ and $\parallel$) is the GL coherence length at zero temperature, and $\Phi_0 = h/2e$ (with $h$ being the Planck's constant) is the flux quantum. The variation of the GL coherence length with temperature can be estimated by~\cite{Tinkham-1996}
\begin{equation}\label{Eq-Coherence}
\xi_i (T) = \frac{\xi_i(0)}{\left(1-T/T_{\rm{c}}\right)^{1/2}}.
\end{equation}
The theoretical prediction of Eq.~(\ref{Eq-Coherence}) for each film is presented in Fig.~\ref{Fig-Coherence}. Both the out-of-plane and in-plane coherence lengths, $\xi_{\perp}(T)$ and $\xi_{\parallel}(T)$, increase with temperature and diverge at $T_c$. At a given temperature below $T_c$, $\xi_{\perp}(T)$ exceeds $\xi_{\parallel}(T)$ in magnitude for each film. However, even at $0.7 T_c$, the out-of-plane coherence length $\xi_{\perp}$  reaches only $\sim$1/7 of the film thickness for the film with the maximum $\xi_{\perp}(0)$.
In the following subsection we therefore explore the resistivity as a function of magnetic field for $T \lesssim 0.7 T_c$, a range in which all the films can be safely regarded as 3D superconductors.

\subsection{Scaling analysis}
For the $x\simeq 1.21$ film, the resistivity vs field curves measured at different temperatures merge into a single horizontal line at high fields, and the situation is the same both for the perpendicular and parallel field cases. Thus, the QGS does not emerge in this film, and we focus our attention on the low-temperature magnetoresistance isotherms of the other ZrN$_ x$ films in the following discussion. Considering the results for the $x\simeq 1.30$, 1.35, 1.42, and 1.50 films are similar, we only present and discuss the results obtained from the two representative films ($x\simeq 1.35$ and 1.50) in detail below.

\begin{figure*}
\includegraphics[scale=0.9]{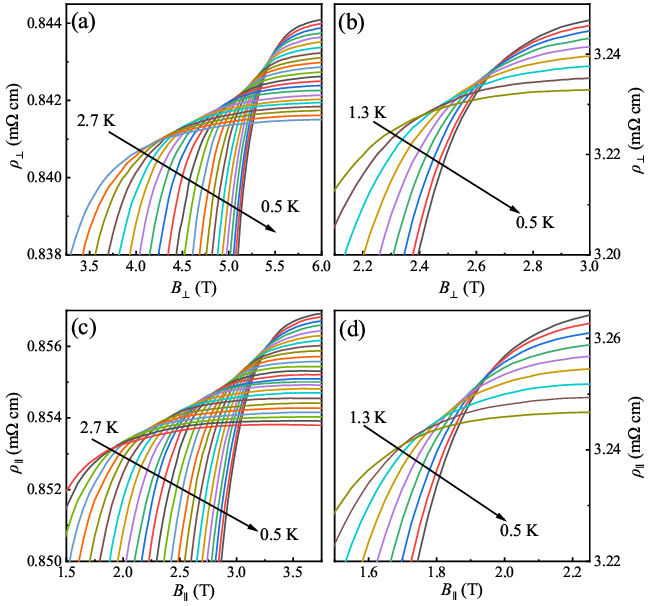}% Here is how to import EPS art
\caption{\label{Fig-RB2}(a) Resistivity $\rho_{\perp}$ vs perpendicular field $B_{\perp}$ measured at temperature from 0.50 to 2.70\,K for the $x\simeq 1.35$ film.  (b) $\rho_{\perp}$ vs $B_{\perp}$ measured at temperature from 0.50 to 1.30\,K for the $x\simeq 1.50$ film. The curves in (c) and (d) are  the $\rho_{\parallel}$ vs $B_{\parallel}$ at different temperatures for the $x\simeq 1.35$ and 1.50 films, respectively. The interval between two adjacent temperatures is 0.10\,K.}
\end{figure*}

Figure~\ref{Fig-RB2}(a) and \ref{Fig-RB2}(b) display the perpendicular field $B_{\rm \perp}$ dependence of the resistivity $\rho_{\rm \perp}$ measured at low temperatures ($T \lesssim 0.7T_c$) for the $x\simeq 1.35$ and $x\simeq 1.50$ films, respectively. Clearly, the magnetoresistance isotherms at low temperatures do not converge into a horizontal line under large perpendicular fields like the behavior of the $x\simeq 1.21$ film or cross at a single point like that of the traditional QPT in 2D superconducting system, but cross at many points located in a relatively large transition area. For $\rho$-$B$ curves measured at parallel field case, the situation is quite similar, which can be seen from  Fig.~\ref{Fig-RB2}(c) (the $x\simeq 1.35$ film) and \ref{Fig-RB2}(d) ($x\simeq 1.50$).

\begin{table*}%[htbp]
\caption{
Relevant fitting parameters of the ZrN$_x$ films with $x\gtrsim1.30$. Here $C_{i}$ ($i=\perp$ and $\parallel$), $B_{\rm ci}^\ast$, and $\nu_i$ are defined in Eq.~(\ref{Eq-actiavated-scaling}). $\tilde{T}_{0i}$ ($i=\perp$ and $\parallel$) is a parameter in Eq.~(\ref{Eq-Infinite-Randomness}) and represents the microscopic temperature scale associated
with the quantum phase transition.}\label{tab-2}
\begin{ruledtabular}
\begin{tabular}{cccccccccc}
Film   & $x$  &  $C_{\perp}$  & $B_{\rm c \perp}^\ast$ & $\nu_\perp$ & $C_\parallel$ & $B_{\rm c \parallel}^\ast$ & $\nu_\parallel$ & $\tilde{T}_{\rm 0 \perp}$ &  $\tilde{T}_{\rm 0 \parallel}$  \\
No. &    &      &  (T)     &     &   &  (T)  &   & (K)  &    (K)     \\
\hline
2 &   1.30 & 0.41 & 5.54 & 1.52 & 0.62 & 3.34 & 1.70 &   5.36 & 6.29 \\
3 &   1.35 & 0.43 & 5.54 & 1.46 & 0.63 & 3.36 & 1.58&   9.36 & 8.98 \\
4 &   1.42 & 0.27 & 3.21 & 1.40 & 0.32 & 2.18 & 1.45&   2.37 & 3.00 \\
5 &   1.50 & 0.20 & 2.70 & 1.64 & 0.22 & 2.01 & 1.72&   2.44& 3.48 \\
\end{tabular}
\end{ruledtabular}
\end{table*}

\begin{figure}[htbp]
\includegraphics[scale=1.08]{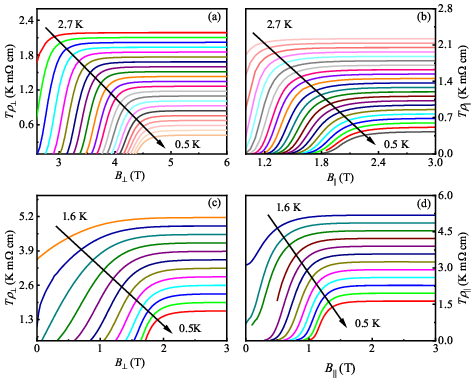}% Here is how to import EPS art
\caption{\label{Fig-TR-B}(a) $T\rho_{\perp}$ vs $B_{\perp}$ and (b) $T\rho_{\parallel}$ vs $B_{\parallel}$ plots for the $x\simeq 1.35$ film at temperatures from 0.50 to 2.70\,K. (c) $T\rho_{\perp}$ vs $B_{\perp}$ and (d) $T\rho_{\parallel}$ vs $B_{\parallel}$ plots for the $x\simeq 1.50$ film at temperatures from 0.50 to 1.30\,K. The temperature interval between adjacent curves is 0.10\,K.}
\end{figure}

For a $d$-dimensional superconductor, the dirty-boson model predicts that the resistivity as a function of field and temperature in the vicinity of the QPT (e.g. superconductor-insulator and superconductor-metal transitions) follows the scaling of the form~\cite{Fisher-1989,Fisher-1990},
\begin{equation}\label{Eq-scaling-dD}
  \rho(T, B)=\rho_c \cdot \left(\frac{T^\ast}{T}\right)^{(d-2)/z}\cdot f\left(\frac{|B-B_c|}{T^{1/z\nu}} \right)
\end{equation}
where $\rho_c$ is the critical resistivity, and $T^{\ast}$ is the characteristic temperature associated with the QPT. Eq.~(\ref{Eq-scaling-dD}) reduces to the power-law scaling form in Eq.~(\ref{Eq-dirtyBoson}) when $d = 2$. For disordered superconductor undergoing superconductor-insulator (or metal) transition, the dynamic exponent $z$ is equal to 1 for all $d$. For $d = 3$, Eq.~(\ref{Eq-scaling-dD}) means that the low-temperature $T \rho$-$B$ data measured at different temperatures all cross at one point, and $T\rho$ vs $|B-B_c|T^{-1/z\nu}$ data in the vicinity of the superconductor-insulator (or metal) transition collapse into two branches for a suited $z\nu$. Figure~\ref{Fig-TR-B} shows the low-temperature $T \rho_i$-$B_i$ ($i=\perp$ and $\parallel$) data for the $x\simeq 1.35$ and $x\simeq 1.50$ films. Clearly, regardless of whether the magnetic field is parallel or perpendicular to the film plane, the $T\rho$-$B$  curves of each film at different temperatures do not intersect near the superconductor–insulator (or metal) transition. Thus, the $\rho(B,T)$ data near the QPT for our films cannot be described by the theoretical prediction of the dirty-boson model for $d = 3$.

\begin{figure}[htbp]
\includegraphics[scale=0.8]{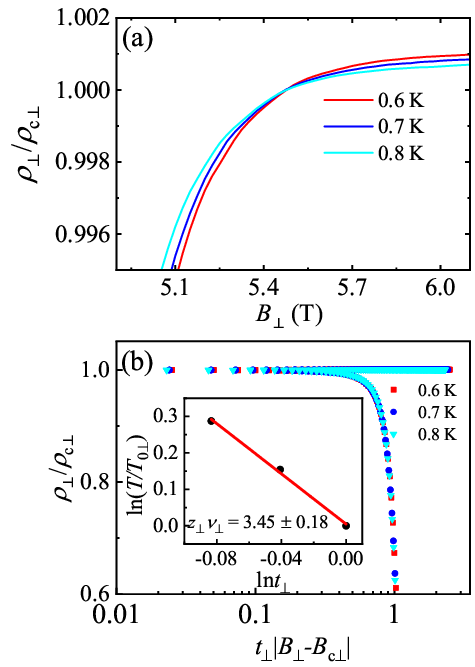}% Here is how to import EPS art
\caption{\label{Fig-zv3}(a) Normalized resistivity $\rho_{\perp}/\rho_{\rm c\perp}$ versus perpendicular magnetic field $B_{\perp}$ measured at 0.60, 0.70, and 0.80 K for the $x\simeq 1.35$ film. Here the values of $\rho_{\rm c\perp}$ and $B_{\rm c\perp}$ are taken to be $\rho_{\rm c\perp}=0.843$\,m$\Omega$\,cm and $B_{\rm c\perp}=5.475$\,T, respectively. (b) Scaling plot of $\rho_{\perp}/\rho_{\rm c\perp}$ versus $t_\perp|B_{\perp} - B_{\rm c\perp}|$ near the critical field $B_{\rm c\perp}$, using data from the three temperatures in (a). The scaling variable $t_\perp$ is defined as $t_\perp = (T/T_0)^{-1/z\nu_\perp}$. Inset: power-law plot of $T$ versus $t_\perp$. The critical exponent $z\nu_\perp$ at $B_{\rm c\perp}$ is extracted from the slope of the linear fit.}
\end{figure}

In fact, the characteristics of the $\rho_i$-$B_i$ ($i=\perp$ and $\parallel$) curves at different temperatures for each ZrN$_{x}$ film in Fig.~\ref{Fig-RB2} closely resemble those of the $R$-$B$ curves under perpendicular magnetic field in 2D superconducting film exhibiting QGS~\cite{Xing-2015,Shen-2016,Xing-2017,Saito-2018,Lewellyn-2019,Zhang-2019,Verzhbitskiy-2020,Huang-2021,Liu Y-2021,Jing-2023}. Thus, we adopt the method for analyzing the QPT in 2D superconductors with QGS to inspect the magnetoresistance isotherms of our films, assuming the $\rho$-$B$ curves at three adjacent temperatures cross at one point and the $\rho(B,T)$ data near the crossing point obey Eq.~(\ref{Eq-dirtyBoson}). For comparative purposes, Eq.~(\ref{Eq-dirtyBoson}) is rewritten as $\rho(B, t)=\rho_{c} f(\delta t^{-1/z\nu})$, where $t = (T/T_0)^{-1/z\nu}$ with $T_0$ being the lowest temperature in the three adjacent temperatures.
As an example, in Fig.~\ref{Fig-zv3} we present the normalized resistivity $\rho_{\perp}/\rho_{c \perp}$ as a function of $t_{\perp}|B_{\perp}-B_{c \perp}|$ measured at 0.60, 0.70, and 0.80\,K in the vicinity of $B_{c \perp}$ for the $x\simeq 1.35$ film. The values of $\rho_{c\perp}$ and $B_{c\perp}$ are taken from the crossing point of the three magnetoresistance isotherms.
By adjusting $t_{\perp}$ for each temperature, we find that the magnetoresistance isotherms collapse onto two branches when the data are plotted as $\rho_{\perp}/\rho_{c \perp}$ vs $t_{\perp}|B_{\perp}-B_{c \perp}|$, as indicated in Fig.~\ref{Fig-zv3}(b). The exponent $z\nu_\perp$ can be obtained directly from a linear fit of $\ln(T/T_0)$ against $\ln t$. Using this method, we obtained $z_\perp\nu_\perp$ ($z_\parallel\nu_{\parallel}$) vs $B_{c \perp}$ ($B_{c \parallel}$) for each film.

\begin{figure*}
\includegraphics[scale=1.0]{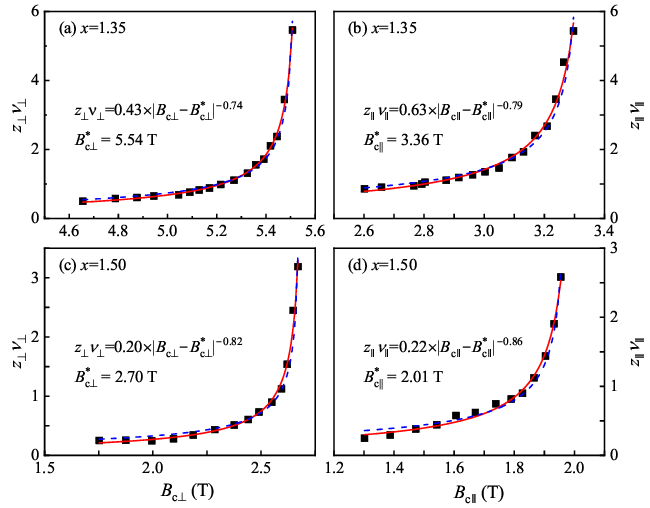}% Here is how to import EPS art
\caption{\label{Fig-zv-B}(a) The critical exponent $z\nu_\perp$ vs the out-of-plane critical field $B_{\mathrm{c}\perp}$ and (b) the critical exponent $z\nu_\parallel$ vs the in-plane critical field $B_{\mathrm{c}\parallel}$ for the $x\simeq 1.35$ film. (c) The critical exponent $z\nu_\perp$ vs the out-of-plane critical field $B_{\mathrm{c}\perp}$ and (d) the critical exponent $z\nu_\parallel$ vs the in-plane critical field $B_{\mathrm{c}\parallel}$ for the $x\simeq 1.50$ film. The solid and dashed curves are the least-squares fits to Eq.~(\ref{Eq-actiavated-scaling}). For the solid curves, the parameter $\nu_i\psi$ is set as an adjustable parameter, while for the dashed ones, $\nu_i\psi$ is fixed at 0.60.}
\end{figure*}

Figure~\ref{Fig-zv-B} presents $z_i\nu_i$ variation with $B_{c i}$ for the $x\simeq 1.35$ and 1.50 films, as indicated. As the temperature is lowered, the coefficient $z_\perp\nu_\perp$ ($z_{\parallel}\nu_{\parallel}$) at first rises gently with increasing  $B_{c\perp}$ ($B_{c\parallel}$). Once $z_i\nu_i$ surpasses $\sim$1, the dependence turns precipitous and appears to diverge as $T\rightarrow 0$\,K. The variation trend in $z_\perp\nu_\perp$ ($z_{\parallel}\nu_{\parallel}$) vs $B_{c\perp}$ ($B_{c\parallel}$) is quite similar to that for $z\nu_\perp$ vs $B_{c\perp}$ in 2D superconductors with QGS. Thus, the $z_i\nu_i$ vs $B_{ci}$ data are least-square fitted to~\cite{Vojta-2009}
\begin{equation}\label{Eq-actiavated-scaling}
  z_i \nu_{i}=C_i|B_{ci}-B^{\ast}_{ci}|^{-\nu_i\psi},
\end{equation}
where $C_i$ is a constant, $B^{\ast}_{ci}$ is the characteristic field as $T \rightarrow 0$, $\psi =0.5$ is the tunneling exponent, $i= \perp$ and $\parallel$ still represent perpendicular and parallel fields, respectively. In the fitting process, $C_i$, $B^{\ast}_{ci}$ and $\nu_i$ are set as adjustable parameters. The fitted results are displayed as solid curves in Fig.~\ref{Fig-zv-B}, and the corresponding parameters are summarized in Table~\ref{tab-2}. From Fig.~\ref{Fig-zv-B}, one can see that both $z_\perp\nu_\perp$ vs $B_{c\perp}$ and $z_\parallel\nu_\parallel$ vs $B_{c\parallel}$ data are well described by the activated scaling law for each film. Both $\nu_\perp\psi$ and $\nu_{\parallel}\psi$ lie between $\sim$0.7 and $\sim$0.8, about 15-35\% greater than that ($\sim$0.61) derived from Monte Carlo simulations in 2D systems~\cite{Vojta-2009}. This level of discrepancy is acceptable for real systems. For comparison, the fitting result with $\nu_i\psi = 0.60$ is also shown as dashed lines in Fig.~\ref{Fig-zv-B}. One can see that the curve obtained with $\nu_i\psi = 0.60$ does not deviate substantially from the experimental data points for each film. It should be noted that the magnetoresistance curves in Fig.~\ref{Fig-RB2} were analyzed using Eq.~(\ref{Eq-scaling-dD}) with $d = 2$. Therefore, the above results indicate that a QGS analogous to that observed in 2D superconducting films occurs in the 3D ZrN$_{x}$ films ($x\gtrsim 1.3$) during the SMT process.

On the other hand, Maestro \emph{et al}.~\cite{Maestro-2010} proposed an activated scaling based on the random transverse field Ising model to describe the conductivity of nanowires near SMT. The activated scaling is also extended to higher dimensions and experimental tested in 2D superconductors with QGS. According to Maestro \emph{et al}., in the vicinity of 0\,K the $R(B,T)$ data can be expressed as~\cite{Lewellyn-2019,Maestro-2010}
\begin{equation}\label{Eq-Infinite-Randomness}
  R\left( \tilde{\delta}, \ln\frac{\tilde{T}_0}{T} \right) =\left(\ln \frac{\tilde{T}_0}{T} \right)^{(d-2)/\psi} \Phi \left( \tilde{\delta}\left[\ln\frac{\tilde{T}_0}{T}\right]^{1/\nu\psi}\right),
\end{equation}
where $\Phi$ is a scaling function, $\tilde{\delta}=|B-\tilde{B}_{\rm c}^{\ast}|/\tilde{B}_{\rm c}^{\ast}$ is the relative distance from the critical field, $\tilde{B}_{\rm c}^{\ast}$ is the critical field of SMT for $T\rightarrow 0$, $\nu$ is the correlation length exponent, $\psi$ is the tunneling exponent, $\tilde{T}_{0}$ is a microscopic temperature scale associated with the quantum phase transition. The prefactor on the right hand side of Eq.~(\ref{Eq-Infinite-Randomness}) become unity for $d=2$. In fact, Eq.~(\ref{Eq-actiavated-scaling}) further indicates that the effective $z\nu$ exponent obtained from the power-law scaling analysis near the crossing points increases with decreasing temperature and finally diverges as $T\rightarrow 0$.
For 2D superconductors with QGS, Lewellyn \emph{et al}. have demonstrated that the variation of the $z\nu$ in Eq.~(\ref{Eq-actiavated-scaling}) with temperature satisfies~\cite{Lewellyn-2019}
\begin{equation}\label{Eq-zv-T}
\frac{1}{z\nu}=\left(\frac{1}{\nu\psi}\right)\frac{1}{\ln(\tilde{T}_0/T)}.
\end{equation}

\begin{figure*}
\includegraphics[scale=1.0]{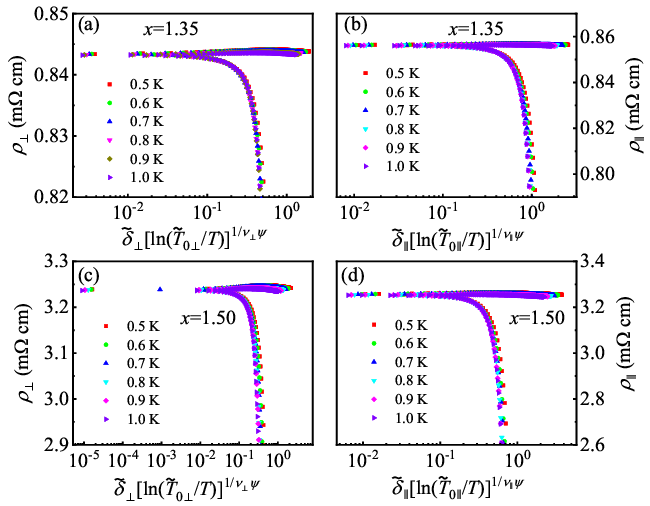}% Here is how to import EPS art
\caption{\label{Fig-actiavatedscaling}The resistivity $\rho_i$ vs the scaling variable $\tilde{\delta}_i[\ln(\tilde{T}_{0i}/T)]^{1/(\nu_i\psi)}$. Panels (a) and (b) correspond to $i=\perp$ and $\parallel$, respectively, for the $x\simeq 1.35$ film, while (c) and (d) correspond to $i=\perp$ and $\parallel$, respectively, for the $x\simeq 1.50$ film.}
\end{figure*}

We compare the low-temperature $\rho_i(B_i,T)$ data of the films near $B_{ci}$ with the prediction of Eq.~(\ref{Eq-Infinite-Randomness}). In the analysis, the values of $\tilde{B}^{\ast}_{ci}$ and $\nu_i$ are set equal to those of $B^{\ast}_{ci}$ and $\nu_i$ obtained from the fits to Eq.~(\ref{Eq-actiavated-scaling}), and the dimensionality $d$ is set to 2. The parameter $\tilde{T}_0$ is determined from the least-squares fits to Eq.~(\ref{Eq-zv-T}). Figure~\ref{Fig-actiavatedscaling} presents the resistivity vs $\tilde{\delta_i}\ln(\tilde{T}_{0i}/T)^{1/\nu_i\psi}$ for the $x=1.35$ and 1.50 films. Clearly, the $\rho_i(B_{i}, T)$ vs $\tilde{\delta}_i\ln(\tilde{T}_{0i}/T)^{1/\nu_i\psi}$ data collapses onto two branches in the vicinity of $\tilde{B}^{\ast}_{ci}$ for each film, indicating that $\rho_i(B_i,T)$ data near the QPT obey the activated scaling law described in Eq.~(\ref{Eq-Infinite-Randomness}). As the magnetic field moves away from the critical field $\tilde{B}_{ci}^{\ast}$, the $\rho_i(B_i, T)$ vs $\tilde{\delta}_i\ln(\tilde{T}_{0i}/T)^{1/\nu_i\psi}$ data gradually deviates from the prediction of Eq.~(\ref{Eq-Infinite-Randomness}); in other words, the activated scaling relation breaks down far from $\tilde{B}_{ci }^{\ast}$. In fact, Eq.~(\ref{Eq-Infinite-Randomness}) is valid only in the vicinity of the quantum critical point ($T\rightarrow 0$\,K). As the temperature increases, corrections to the scaling relation due to temperature  must be taken into account~\cite{Lewellyn-2019}. Recently, Cui \emph{et al}.~\cite{Cui-2023} reanalyzed the magnetoresistance isotherms of the thri-layer Ga films in Ref.~\cite{Xing-2015} and LaAlO$_3$/SrTiO$_3$ interface superconductors in Ref.~\cite{Shen-2016} using Eq.~(\ref{Eq-Infinite-Randomness}) by taking the irrelevant corrections into account, and found that these magnetoresistance data can be well described by the corrected activated scaling relation.

In this subsection, the magnetoresistance isotherms were first analyzed using the power-law scaling Eq~(\ref{Eq-scaling-dD}) with $d=2$. It is found that the critical exponents $z_{\perp}\nu_{\perp}$ and $z_{\parallel}\nu_{\parallel}$ grow monotonically and eventually diverge as $T\rightarrow 0$\,K. At the same time, the same data near the critical field can be described by the activated scaling Eq.~(\ref{Eq-Infinite-Randomness}) with $d=2$, which applies to 2D superconductors exhibiting QGS. Thus, these 3D ZrN films exhibit behaviors nearly identical to those of 2D superconductors with QGS during the superconductor-insulator (or metal) transition. In this sense, the 3D ZrN films display 2D superconducting characteristics and exhibit QGS during the QPT.

\subsection{Discussion}
Theoretically, the emergence of rare regions due to quenched disorder is identified as the principal origin of quantum Griffiths singularities, independent of whether the system is two- or three-dimensional~\cite{Maestro-2010,Vojta-2016,Vojta-2019}. In ZrN$_x$ epitaxial films deposited by reactive sputtering method, point defects (such as Zr vacancies and N interstitials ) and dislocations are unavoidable. These intrinsic defects are the primary origin of quenched disorder in ZrN films. It has been established that disorder in 2D superconductors drives the superconducting order parameter into a spatially inhomogeneous state~\cite{Ghosal-2001,Skvortsov-2005,Sacepe-2008}. Recently, it is found that similar situation also emerges in some 3D superconductors~\cite{Kamlapure-2013,Parra-2021}, including certain metal-nitride compounds~\cite{Kamlapure-2013}. This inhomogeneity behaves as the appearance of superconducting rare regions (superconducting puddles or droplets) in the sample. When subjected to a sufficiently strong field at low temperatures, the sample undergos a superconductor to insulator (or metal) transition. In the vicinity of the transition, the slow dynamics of the superconducting rare regions gives rise to QGS.

A puzzling aspect of our results is why 3D ZrN$_x$ films exhibit the same scaling behavior as 2D films with QGS during the superconductor-to-insulator (or metal) transition. In 2012, Giraldo-Gallo \emph{et al.}~\cite{Giraldo-2012} reported that the low-temperature magnetoresistance isotherms of BaPb$_{1-x}$Bi$_x$O$_3$ ($x=0.24$ and $0.25$) single crystals cross at a single point and follow the power-law scaling form of a 2D system, i.e., Eq.~(\ref{Eq-dirtyBoson}). They attribute their observations to a hidden two-dimensional character originating from the nanostructure of the BaPb$_{1-x}$Bi$_x$O$_3$ samples, and suggest that it manifests as a percolating network of superconducting sheets. This scenario could also be valid in the 3D ZrN$_x$ films. As mentioned above, the disorder makes the superconductivity inhomogeneity in the ZrN$_x$ films. The superconductor to insulator (or metal) transition, therefore, can be reasonably understood in framework of the percolation model. Near the quantum critical point, the dynamics of the superconducting rare regions dominate the transition behavior. These rare regions may exhibit effective 2D characteristics near the transition, leading to the emergence of 2D QGS in these 3D ZrN$_x$ superconducting films. According to Spivak \emph{et al.}~\cite{Spivak-2008,Kapitulnic-2019}, the dissipation in superconductors scales with the surface of the rare region rather than its volume in the limit of large rare-region size. This illustrates the validity of the above physical picture from another perspective. Direct observation of the dynamics of superconducting rare regions remains an open challenge in both 3D and 2D superconductors.

\section{Conclusion}\label{secIV}

In summary, we have systematically investigated the low-temperature electrical transport properties of $\sim$200-nm-thick epitaxial ZrN$_x$ films with varying nitrogen compositions. All films reveal 3D superconductivity characteristics below their superconducting transition temperatures. Both out-of-plane and in-plane magnetic fields can drive the films to transition from a superconducting state to an insulating (or bad-metal) state. For films with $x\gtrsim 1.30$, the low-temperature magnetoresistance isotherms under out-of-plane and in-plane fields cross over a broad field range rather than at a single point, analogous to that for 2D superconducting systems with QGS under out-of-plane fields. Although the films are 3D with respect to superconductivity, the magnetoresistance isotherms at adjacent temperatures follow the power-law scaling predicted for 2D superconducting systems, rather than that for 3D systems. The effective critical exponent $z\nu$, extracted from the magnetoresistance isotherms using the 2D power-law scaling, increases with decreasing temperature and diverges as the quantum phase transition is approached. In addition, the resistivity data near the superconductor-insulator (or metal) transition obey an activated scaling form describing the quantum phase transition in 2D superconducting systems. Our results clearly indicate that the QGS occurring in 2D superconducting systems appears in these 3D ZrN$_x$ ($x\gtrsim 1.30$) films. The rare regions in these 3D ZrN$_x$ films are suggested to be 2D, which causes the emergence of 2D QGS therein.

\begin{acknowledgments}
This work is supported by the National Natural Science Foundation of China (Grant No. 12174282) and Frontier Fundamental Research Program of Tianjin University (Grant No. 2025XJ21-0005).
\end{acknowledgments}

\end{document}